\documentclass[a4paper,fleqn]{cas-dc}
\usepackage[numbers,sort&compress]{natbib}

\begin{document}
\let\WriteBookmarks\relax
\def\floatpagepagefraction{1}
\def\textpagefraction{.001}
\def\citedash{--} 
\shorttitle{Investigation of Widefield NV-Center Magnetic Imaging for Non-Destructive Materials Testing}
\shortauthors{N.Mathes et~al.}

\title [mode = title]{Investigation of Widefield NV-Center Magnetic Imaging for Non-Destructive Materials Testing}

\author[1]{N. Mathes} [orcid = 0000-0001-5476-9470]

\author[2]{M. Feuerhelm}[orcid = 0009-0001-4176-9516]

\author[2]{S. Philipp}[orcid = 0000-0001-5890-035X]

\author[3]{I. Soldatov}[orcid =  0000-0002-2911-1842]

\author[1]{P. D'Astolfo}[orcid = 0009-0004-3584-9992]

\author[1]{P. Knittel}[orcid=0000-0002-2769-3615]

\author[2]{T. Straub}[orcid = 0000-0002-6874-7675]

\author[1]{J. Jeske}[orcid = 0000-0003-3532-506X]
\cormark[1]
\ead{jan.jeske@iaf.fraunhofer.de}

\author[1]{R. Quay}[orcid = 0000-0002-3003-0134]

\author[1,4]{X. Vidal} [orcid = 0000-0001-8026-2016]
\cormark[1]
\ead{xavier.vidal@tecnalia.com}

\cortext[1]{Corresponding authors}

\affiliation[1]{organization={Fraunhofer Institute of Applied Solid State Physics IAF},
                addressline={Tullastraße 72}, 
                city={Freiburg},
                postcode={79108}, 
                country={Germany}}

\affiliation[2]{organization={Fraunhofer Institute for Mechanics of Materials IWM},
                addressline={Wöhlerstraße 11}, 
                postcode={79108}, 
                city={Freiburg},
                country={Germany}}

\affiliation[3]{organization={Leibniz Institute for Solid State and Materials Research Dresden (IFW Dresden)},
                addressline={Helmholzstraße 20}, 
                postcode={01069}, 
                city={Dresden},
                country={Germany}}

\affiliation[4]{organization={TECNALIA, Basque Research and Technology Alliance (BRTA)},
                addressline={Astondo Bidea 700}, 
                postcode={48160}, 
                city={Derio},
                country={Spain}}

\begin{abstract}
Due to progressing miniaturization, material fatigue can be expected to gain importance on microscopic scales. Nevertheless, as of today, most techniques for non-destructive testing are optimized for macroscopic scales. Therefore, new approaches that are applicable to miniaturized samples need to be explored. High spatial resolution imaging of a material's magnetic stray field allows for a sensitive detection of micro-structural changes due to the direct local interplay of magnetic material properties, strain and defects. Up to now, this approach has rarely been used for materials testing since high spatial resolution magnetic sensing traditionally is a very challenging process. Novel quantum sensing techniques enable new sensing approaches that offer the potential to close this gap. One of the most prominent quantum sensors, the nitrogen vacancy center in diamond, is investigated for non-destructive testing applications within the scope of this work. An experimental system based on a widefield sensing approach was developed to image magnetic stray field distributions within seconds to minutes in an area of up to 1 x 1$\,$mm². Magnetic sensitivities below 10$\,$µT/$\sqrt{\text{Hz}}$ and a spatial resolution between 1 and 2 µm could be reached. The technique offers a high mechanical stability which is crucial for non-destructive testing applications. Comparing with magneto-optical Kerr effect measurements it could be shown that captured magnetic field maps are strongly correlated to the magnetic material's surface domains but contain additional information from deeper inside the sample. Characteristic changes of the magnetic stray field distribution could be detected for an electrical steel sample after cyclic loading. A potential marker for early fatigue damage was deduced by evaluating the splitting gradient distribution, while an evaluation using 2D Fourier transforms can provide additional insight.
\end{abstract}


\maketitle

\section{Introduction}
Due to its very complicated and statistical nature, material fatigue is not yet fully understood and remains responsible for most material failure incidents to date \cite{lavenstein2019micro}. The degradation process takes place gradually on a microscopic scale before any macroscopic damage, such as visible cracks or bending, becomes apparent \cite{Suresh_1998}. Most non-destructive testing (NDT) approaches were developed for rather large material parts and rely on the detection of cracks and material discontinuities. Due to progressing miniaturization, material properties on small scales can be expected to become increasingly relevant in fields such as medical technology or micro-electro-mechanical systems (MEMS). To enable an early detection of material fatigue in such components, but also to gain a better understanding of the underlying microscopic processes, alternative NDT methods on small scales are required.

There are many different NDT approaches, such as: microscopic  inspection, X-ray tomography, acoustic techniques or eddy current testing \cite{Arnold_2023}. For materials possessing a magnetic order, such as many types of steel, it can be especially insightful to exploit local magnetic properties to assess the material's state. The coupling of magnetization, microstructure of the host crystal and local stress distributions is a fundamental feature of ferromagnetic materials \cite{kronmuller2003micromagnetism}. Established magnetic non-destructive testing techniques that utilize this relation, such as the metal magnetic memory method \cite{MMM_review, mmm_chen_research_2017, WANG2010513} and the detection of magnetic flux leakage \cite{flux}, are widely used in practice. \cite{Jiles1988_MagNDEReview1, Jiles1990_MagNDEReview2}. At macroscopic scales (hundreds of micrometers to millimeters), Magneto-Optical Indicator Films (MOIF) can be utilized for the real-time mapping of magnetic stray fields, providing an immediate visual representation of defects \cite{MOIF1, MOIF2} or domain structures \cite{SCHAFER2019221}. Nevertheless, these techniques are either impractical or fundamentally limited when applied to miniaturized samples due to requirements of strong magnetization, excitation fields or large sensing volume \cite{Wang_2012_revNDT}.

Fatigue-induced changes of the microstructure can be detected by high spatial resolution imaging of the magnetic stray field emitted by the sample. This approach remains rarely applied for NDT to date, although some research can be found \cite{stegemann}. Traditional scanning probe approaches are hardly applicable for NDT due to their mechanical delicacy, small field of view and long measurement times. Measuring the magneto-optical Kerr effect (MOKE) allows to image the surface magnetization of a sample with a high spatial resolution, but the method is very hard to quantify and only gives a very limited information about magnetic microstructure of the sample \cite{Schaefer_1998}.

Novel quantum sensing approaches provide new opportunities for high spatial resolution magnetic sensing. Compared to classical sensors, quantum sensors are based on the measurement of individual quantum states \cite{degen_quantum_2017}. The NV center in diamond is one of the most promising candidates for nano- and microscale quantum sensing \cite{doherty_nitrogen-vacancy_2013, abe_tutorial_2018, schirhagl_nitrogen-vacancy_2014}. It consists of a substitutional nitrogen atom next to a vacancy in the diamond crystal lattice. This configuration forms a localized electronic system with a spin of one in the ground state. The unique properties of the diamond host crystal, together with the fluorescent nature of the NV center, allow initialization, coherent manipulation, and optical readout of the electronic spin state. Within the scope of this work, a widefield sensing approach is used providing a favorable combination of spatial resolution, field of view, sensitivity and measurement time well suited to measure magnetic material stray fields \cite{levine_principles_2019, MATHES2023247, Rondin2014_NVMagnetometryReview,Scholten2021_WidefieldNV}. A thin layer of NV centers near the surface of a diamond plate is used for sensing. The fluorescence intensity distribution emitted from the layer, containing direct information about the local spin state of the NV centers, is imaged on a camera sensor. Using this approach, quantum sensing can be performed with a large field of view and a diffraction limited spatial resolution. Due to the parallel readout of the camera sensor, scanning can be avoided allowing for short measurement times. The technique is mechanically robust and does not require any magnetic shielding, offering significant potential for industrial applications.

First attempts to image stress distributions and material damage have already been carried in a scanning configuration at the millimeter scale \cite{Zhou2021_DamageSteelNV}. NV-center quantum sensing has been used in several configurations on the nano scale \cite{Xu2023_RecentAdvances}. Some literature can be found exploring NV center quantum sensing for NDT, limited to special cases or with restricted imaging resolution \cite{grindingburns, Xiang_2025, nvfluxleakage}.

This work explores the applicability of widefield NV center quantum sensing for non-destructive testing. An experimental setup was developed and characterized, providing high flexibility and ease of use for different samples and measurement approaches. A variety of measurements on different material samples were carried out. The measurement results demonstrate a strong potential for the application of widefield NV center quantum sensing to NDT applications. The physical background of the measurement technique is discussed in detail and complemented by simulations. Two different methods for evaluating the measurements for non-destructive testing are proposed. 

\section{Magnetic Investigation of Fatigue Damage on the Mesoscale}

\subsection{Impact of Fatigue Mechanisms on the Magnetic Stray Field}

A macroscopic ferromagnetic material generally develops a complicated, three dimensional magnetic domain structure, also called magnetic microstructure, minimizing the total free energy consisting of various competing contributions \cite{Schaefer_1998}. The magnetic microstructure is directly related to local material properties, such as magnetic anisotropy, defect concentrations and strain \cite{kong}, but also to the geometrical shape of the specimen. Strain affects the magnetic microstructure via inverse magnetostriction, also called the Villari effect \cite{magnetostriction, ives2023toward}. Flux closure structures minimizing magnetic stray fields are usually energetically favorable. However, material imperfections prevent complete compensation of stray fields.

\begin{figure}
    \centering
    \includegraphics[width=1\linewidth]{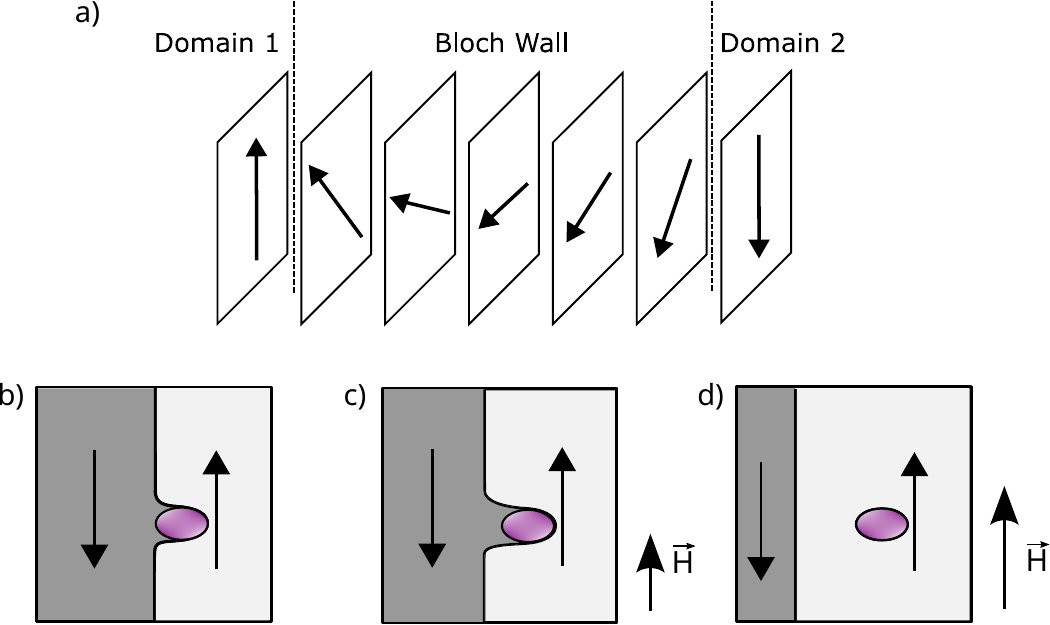}
    \caption{a) Schematic illustration of a domain wall separating two oppositely magnetized domains. The wall involves a gradual rotation of the magnetization between the domains. The typical width of a Bloch wall in bulk ferromagnetic materials is in the order of several tens to hundred nanometers b,c) Sketch of domain wall pinning at a material defect (purple). Although further wall motion would reduce magnetic free energy, the defect locally increases the energy barrier and hinders movement. d) The pinning barrier can be overcome by applying a sufficiently strong external magnetic field.}
    \label{fig:domain_walls}
\end{figure}

All fatigue-induced processes generate localized stress fields and structural defects that modify the magnetic microstructure via magnetoelastic coupling. In particular, stress-induced changes in magnetoelastic energy influence domain wall motion and domain configuration. Since magnetic stray fields can be expected to  predominantly arise from domain walls (see Fig. \ref{fig:domain_walls}a) and spatial variations of magnetization near the surface, alterations of the magnetic microstructure are expected to produce measurable changes in the stray field distribution \cite{DONG2009323, LIHONG2008184, strain_magnetic_signal}.

Domain wall pinning (see Fig. \ref{fig:domain_walls}b-d) plays a crucial role in the evolution of the magnetic microstructure. Domain walls represent regions of rapid magnetization rotation and therefore carry finite energy. In an ideal defect-free material, domain walls would move towards energetically favorable positions determined by the minimization of the total magnetic free energy. However, material defects such as dislocations, inclusions, or precipitates locally modify the magnetic energy landscape and can hinder wall motion. As a result, domain walls may become pinned at defect sites, leading to irregular domain configurations and corresponding variations in the magnetic stray field distribution. The energy barriers introduced by such defects can be overcome by sufficiently strong external magnetic fields or applied stress, resulting in irreversible wall motion. Domain wall pinning therefore strongly influences magnetic hysteresis behavior and contributes substantially to how the magnetic microstructure evolves under cyclic loading. Since fatigue processes introduce and accumulate microscopic defects, an increase in pinning centers is expected, which can significantly modify domain wall mobility and, consequently, the observed magnetic stray field distribution.

Continuous cyclic loading of a material causes the formation of dislocations in the crystal lattice. These dislocations represent localized disruptions in the periodic arrangement of atoms and are introduced when the material is subjected to repeated stress cycles. As cyclic loading continues, dislocations begin to interact and accumulate, forming dislocation pile-ups at obstacles such as grain boundaries, precipitates, or other microstructural features. These pile-ups create localized stress fields within the material that can significantly alter local material properties. Due to the magnetoelastic effect, these stress fields act as energy barriers for domain wall movement, effectively pinning domain walls in their vicinity and preventing them from reaching their energetically most favorable positions. \cite{DONG2009323, LIHONG2008184}

Further cyclic loading causes dislocations to organize into persistent slip bands (PSBs), which represent localized regions of intense plastic deformation. PSBs manifest at the material surface as extrusions and intrusions, surface irregularities where material is either pushed out or drawn inward along slip planes \cite{ lavenstein2019micro,Suresh_1998}. Additionally, void formation can occur within the material bulk due to vacancy condensation and dislocation interactions. \cite{Suresh_1998} These microstructural features, including PSBs, extrusions, intrusions, and voids, cause strong pinning of domain walls due to the significant local stress fields they generate, and therefore lead to a marked increase in the stray fields in these areas. The altered domain wall configurations forced by this pinning result in locally increased magnetic stray field amplitudes that can be detected by high-resolution magnetic imaging.

The increased stress concentration around PSBs, extrusions, intrusions, and voids act as a preferential starting point for microcrack nucleation and growth. These regions represent weak points in the material where the local stress state exceeds the material's resistance to crack initiation. Microcracks act as discontinuities in the material, disrupting the magnetic domain structure and leading to strong formation of stray fields at their boundaries. The sharp geometrical discontinuity introduced by a microcrack forces domain walls into unfavorable configurations, generating pronounced magnetic stray field patterns.

Further cycling causes the microcracks to grow and coalesce, eventually leading to the formation of a dominant crack and ultimately to material failure. Throughout this progression, the magnetic stray field distribution evolves characteristically, reflecting the underlying damage state of the material.

Therefore, magnetic stray field measurements allow us to follow the entire evolution of fatigue damage, starting from the formation of dislocations and dislocation pile-ups, progressing to PSBs with associated surface features and internal voids, advancing to microcrack nucleation and growth, and ultimately leading to macroscopic crack formation and material failure. Looking at the stray field distribution during this process, we expect an initial increase in small stray field patterns stemming from dislocation formation distributed throughout the material, leading to widespread but subtle domain wall pinning effects. When PSBs and microcracks start to form, we would expect few but strong, localized stray field patterns to emerge, correlating directly to the positions of the incurring material damage. This characteristic evolution from distributed weak signals to localized strong signals provides a potential marker for non-destructive early fatigue damage detection.

The correlation between the measured magnetic stray field distribution and the underlying magnetic domain structure is not straightforward. While the magnetic field generated by a magnetization distribution can be calculated unambiguously by solving Maxwell's equations in the magnetostatic limit \cite{engel2005calculation, Jackson, Nolting3}, the inverse process is generally ill-posed and does not produce a unique solution. Different magnetization configurations can produce identical external stray field distributions due to the non-uniqueness of the corresponding scalar magnetic potential. Consequently, an exact reconstruction of the three-dimensional domain structure from stray field measurements alone is generally not possible. Quantitative determination of the magnetization distribution requires additional assumptions or constraints, such as reduced dimensionality or known sample geometry, as is often the case for thin magnetic films. Nevertheless, high spatial resolution imaging of the stray field distribution provides valuable information about changes in the micromagnetic configuration and allows the detection of defect- and stress-induced magnetic variations within the sample.

\subsection{Magneto-mechanical testing of mesoscale samples} 
As already mentioned, non-destructive testing of miniaturized material samples is very challenging, since traditional non-destructive testing approaches are only applicable for macroscopic material parts. Optically-pumped magnetometers (OPM) \cite{Tierney2019_OPM_Review}, which are quantum sensors based on an atomic vapor cell, have recently been used to investigate the relationship between the stress-strain dependence in fatigue testing and the relationship between the magnetic field generated by the inverse magnetostrictive effect (or Villari effect) \cite{Bao2010_PiezomagneticHysteresis,Bao2011_FatigueMagneticMechanical,Bao2012_MagnetomechanicalFatigue,Riesgo2020_VillariLowStrain} using material samples with dimensions in the order of several hundred microns \cite{Koss2022_OPM_FatigueSteel,Thiemann2022_OPM_InitialDamage}. The results show first steps toward an understanding of fatigue damage related changes in the magnetic signal as a function of a cyclic loading. The high sensitivity of the OPM comes at cost of spatial resolution, resulting in the measurement of one component of the magnetic stray field at a time, at a relatively large distance from the sample surface. A more detailed understanding of the formation of damage sites and the corresponding interaction with the magnetization and the corresponding stray field requires a spatially resolved measurement of the local stray field distribution close to the surface. Spatially resolved NV center quantum sensing offers the potential to close this gap.

\section{Widefield NV-Center Magnetometry for Magnetic Materials Testing}
\subsection{NV-Center Magnetometry}
\begin{figure}
\centering
  \includegraphics[width=1\linewidth]{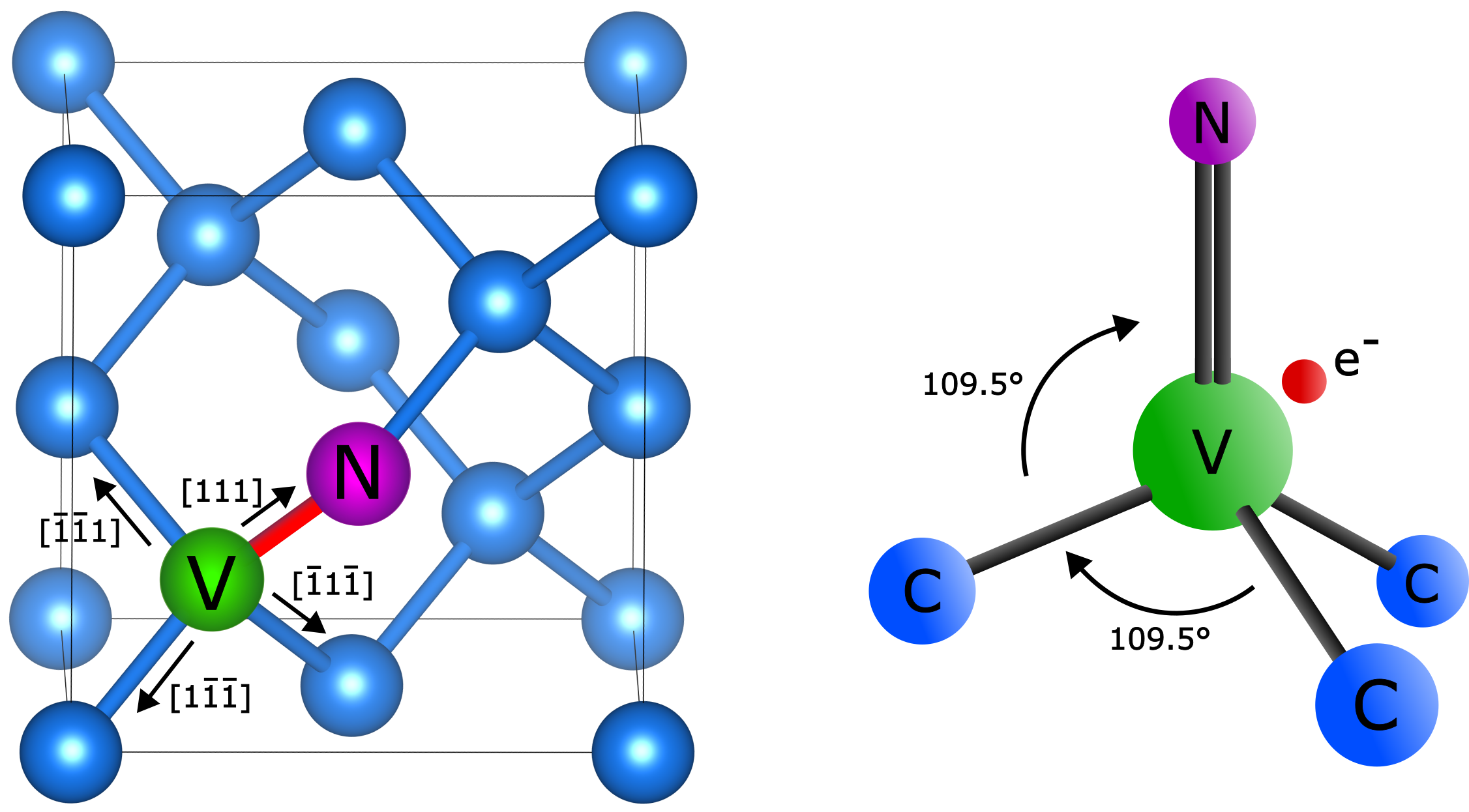}
  \caption{The basic structure of the NV center in diamond consisting of a substitutional nitrogen atom and a vacancy (V) within a diamond lattice consisting of carbon atoms. For the NV$^{-}$-center, an additional electron is trapped in the vacancy.}
  \label{fig:structure}
\end{figure}
The NV center in diamond is a fluorescent color center with unique characteristics allowing for its use as an atomically small quantum sensor \cite{doherty_nitrogen-vacancy_2013, schirhagl_nitrogen-vacancy_2014}. It consists of a substitutional nitrogen atom next to a vacancy (see Fig. \ref{fig:structure}) in the diamond crystal lattice. This configuration leads to the formation of an electronic system with spin 1 which can be excited by a laser in the wavelength range of 515 nm to 565 nm. It then relaxes back to the ground state via emission of a red shifted photon or by an alternative non-radiative inter system crossing (ISC) path. 
The fluorescence zero phonon line is located at 638$\,$nm.  There are four different orientations of NV centers within the diamond crystal lattice in the directions $[111]$, $[1\bar{1}\bar{1}]$, $[\bar{1}\bar{1}1]$ and $[\bar{1}1\bar{1}]$.

\begin{figure}
 \centering
 \includegraphics[width=1\linewidth]{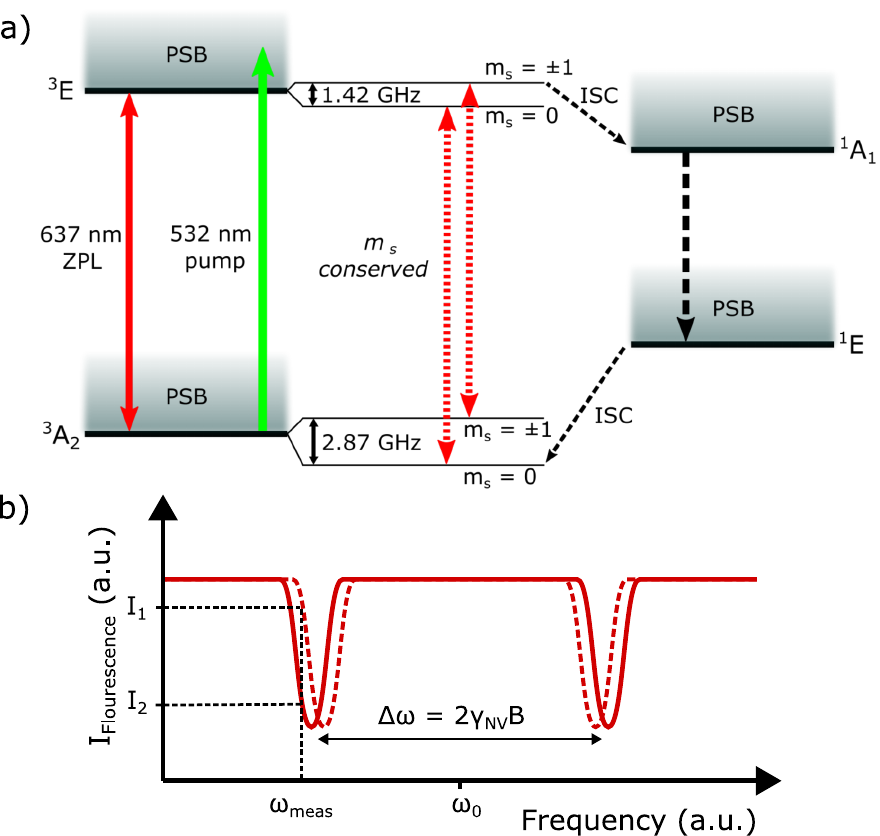}
 \caption{a) Simplified Energy Diagram of the NV$^{-}$-center. The left hand side shows the triplet states and right hand side the singlet states. The fluorescence zero phonon line (ZPL) wavelength is 637$\,$nm, the spin state is conserved for the fluorescent transitions. Absorption of the 532$\,$nm laser light results in an excitation of the electronic system from the ground state to the excited state. The transitions in most cases involve the phononic sideband (PSB). The zero field splitting of the $m_s$ =  0 and $m_s$ = $\pm$1 spin ground states is 2.87$\,$GHz (1.42$\,$GHz for the excited state). An alternative inter system crossing (ISC) decay path via an excited singlet results in a spin dependency of the fluorescence emission intensity. The ISC is more likely for the $m_s$ = $\pm$1 spin states and almost exclusively ends up in the $m_s$ =  0 spin ground state enabling to spin polarize the NV center. Transitions between the spin states are performed via resonant microwave radiation. b) Sketch of the ODMR spectrum when applying a magnetic field. The resonances corresponding to the transitions from m$_s$ = 0 to m$_s$ = $\pm$1 are separated due to the Zeeman splitting. Fixing the microwave frequency $\omega_{\text{meas}}$ at the highest slope position of one of the ODMR resonance peaks allows to detect a change of fluorescence intensity when there is a displacement of the peak position. }
 \label{fig:energy_levels}
\end{figure}

A simplified energy diagram of the NV center electronic system is presented in Fig. \ref{fig:energy_levels}a. At room temperature, the transitions mostly involve the phononic sideband (PSB). The zero field splitting of the $m_s$ = 0 and $m_s$ = $\pm$1 spin states is 2.87$\,$GHz at room temperature. Due to an alternative inter system crossing (ISC) decay path more likely to occur for the $m_s$ = $\pm$1 spin quantum state, the spin state can be determined by measuring the fluorescence intensity. The $m_s$ = $\pm$1 spin states are called the dark states and the $m_s$ =  0 spin state is called the bright state. The NV center spin state can be initialized by laser excitation since the ISC path leads to a depopulation of the $m_s$ =  1 spin states. Spin transitions between the $m_s$ =  0 and the $m_s$ =  $\pm$1 spin states are controlled using resonant microwaves. Sweeping the microwave frequency while exciting the NV center with a laser allows to determine the spin state optically by measuring the fluorescence intensity. This technique is called optically detected magnetic resonance (ODMR) \cite{abe_tutorial_2018}. Due to the Zeeman effect, the splitting of the $m_s$ =  $\pm$1 spin states is approximately linearly dependent on the local magnetic field.

The exceptional properties of the diamond host crystal provide a very long spin coherence time at room temperature allowing for a controlled manipulation and readout of the spin quantum state. In the presence of small external magnetic fields the Hamiltonian of the NV centers can be approximated to: \cite{doherty_nitrogen-vacancy_2013}:
\begin{equation}
H = D S_z^2 + \gamma \vec{B} \cdot \vec{S}.
\end{equation}
$D$ quantifies the the zero-field splitting and $S_z$ is the spin z operator. The second term arises from the interaction of the external magnetic field with the NV center electronic spin. $\vec{S}$ is the spin operator, $\gamma$ is the gyromagnetic ratio of the NV center and $\vec{B}$ is the external magnetic field. For small magnetic fields, only the magnetic field component aligned with the NV axis leads to an effect due to the symmetry of the NV center electronic states. Therefore, the Zeeman contribution to the spin energy levels is:
\begin{equation}
\Delta E_{Zeeman} \approx \frac{2 g_e \mu_B}{h}B_z.
\end{equation}
$\mu_B$ is the Bohr magneton, $h$ is the Planck constant, $B_z$ is the magnetic field in z direction and $g_e$ is the gyromagnetic ratio of the electron.

The magnetic sensitivity is defined as the smallest detectable change of magnetic field. It is usually normalized to unit time for being able to compare measurements with different integration times. The normalized shot-noise limited magnetic sensitivity $\eta$ can be estimated from the measured ODMR spectrum \cite{sensitivity, sensitivity2}: 
\begin{equation}
\label{eq:sensitivity}
\eta \approx \frac{4h}{3 \sqrt{3} g_e \mu_B} \frac{\Delta\omega}{C \sqrt{R_0}} .
\end{equation}
$C$ is the ODMR contrast, $\Delta\omega$ is the FWHM resonance linewidth, $R_0$ is the baseline detection rate. The detection rate strongly depends on the number of NVs in the sensing volume. Therefore, for a fixed optical path, a higher number of NVs generally increases sensitivity.

Alternatively, the microwave frequency can be fixed at the highest slope position of one of the ODMR resonances. This technique allows for a vastly reduced measurement time, but the detectable magnetic field range is also limited by the peak width (see Fig. \ref{fig:energy_levels}b). 
\subsection{Experimental Setup}
\begin{figure}
\includegraphics[width=0.85\linewidth]{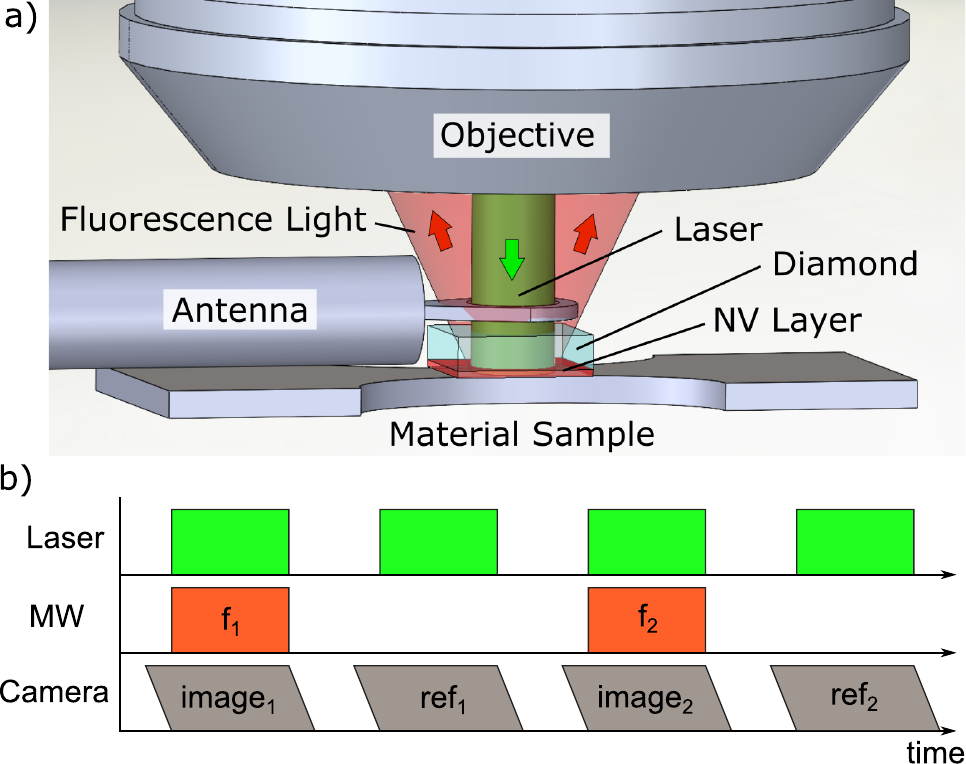}
\centering
\caption{\label{fig:setup} a) Simplified sketch of the widefield magnetometry setup. The NV-layer located near the top surface of the diamond is illuminated from the bottom using a collimated 532$\,$nm laser beam. The fluorescence light is collected by the objective and imaged onto a camera sensor. The dichroic mirror reflects the green laser light letting only the fluorescence light pass. The AOM is needed for pulsing the laser, the image data is captured by a camera sensor. The presented graphic contains components from \cite{componentlibrary}. b) Scheme of the used cw-ODMR protocol. The microwave frequency is swept and an additional reference image is captured for each frequency step. The angled sides of the camera frames illustrate the rolling shutter of the camera. Laser and microwaves are only switched on when all camera pixels are exposed in order to minimize rolling shutter effects.}
\end{figure}

The measurements presented in this work were carried out in a home-built widefield NV center magnetometry setup (see scheme in Fig. \ref{fig:setup}a). The same setup is used as presented and characterized in \cite{MATHES2023247}. For the measurements presented in this work, a long working distance air objective was used allowing to place a loop antenna in between the microscope objective and the diamond. In order to excite the NV layer in a wide area, a collimated 532$\,$nm laser beam is focused on the back focal plane of a 20x magnification and 0.45$\,$NA objective resulting in a collimated laser beam. The maximum field of view is limited by the Gaussian laser intensity distribution to a diameter of approximately 300$\,\mu$m. The laser power focused on the back focal plane of the objective is approximately 100$\,$mW. Since the NV center fluorescence light is collected through the diamond plate, the effective numerical aperture is reduced. Assuming a refractive index of 2.4 for the diamond plate, 1.5 for the glass plate, the effective numerical aperture was calculated to be 0.19. Due to limited space, a tube lens with a focal length slightly larger than the objective design tube lens (200$\,$mm instead of 180$\,$mm) is used and the effective magnification is 18x. The diffraction limited spatial resolution can be estimated by the width of the Airy disk  \cite{kotlyar2010modeling}:
\begin{equation}
d_{Airy, FWHM} \approx \frac{0.51 \cdot \lambda}{\textrm{NA}} = \frac{0.51 \cdot 680\,\text{nm}}{0.19}\approx 1.82\,\text{µm}
\label{eq:resolution}
\end{equation}
Using test samples, the magnetic imaging spatial resolution was determined experimentally to be smaller than 2 µm and the magnetic sensitivity to be below 10 µT/$\sqrt{\text{Hz}}$.

The laser beam is switched on and off using an acousto-optic modulator (AOM), the switching time is about 12$\,$ns. The backward fluorescence light emitted by the NV centers is collected by the same microscope objective and imaged on a sCMOS camera sensor. In order to match the pixel size to the diffraction limit, a 2x2 pixel binning was used resulting in a sample area of about 720$\,$nm$^2$ per pixel. A loop antenna is used to generate a homogeneous microwave distribution in the measurement area.

Magnetic maps are generated by measuring the ODMR spectrum for each pixel in parallel using a continuous wave (cw) ODMR protocol (see Fig. \ref{fig:setup}b). Laser and microwaves are only active when all pixels are exposed in order to reduce effects arising from the rolling shutter of the camera. The camera exposition time is about 20$\,$ ms. This value is optimized for each measurement due to differences in sample reflectivity or interference effects. A frequency range of 170$\,$MHz from 2.835$\,$GHz to 2.905$\,$GHz is scanned with steps of 500$\,$kHz. For each frequency step an additional reference image without microwave radiation is recorded. Each measurement image is divided by the respective reference image in order to cancel out low frequency noise sources such as drifts or laser power deviations. The data collected from 30 frequency sweeps is averaged in order to increase the signal-to-noise ratio. The measurement result is a full ODMR curve for each camera pixel. The final magnetic field image is obtained by fitting a Lorentzian to each ODMR resonance for each pixel and subtracting the resonance peak positions. 

The measurement performance strongly depends on the quality of the diamond hosting the NV layer. Crystal quality and abundance of impurities and other non-zero spin defects (such as $^{13}$C, $^{14}$N and $^{15}$N) play a crucial role as well as the NV density in the layer. The diamond plate used for the presented  measurements is based on a commercial electronic grade type IIa diamond with a (100) surface orientation. The diamond of size 4 x 4 x 0.5 mm$^3$ was CVD homoepitaxially overgrown at Fraunhofer IAF in order to produce a well defined and homogeneous NV layer near the diamond surface. During CVD overgrowth, a N/C ratio of 20000$\,$ppm was used leading to a P1 center concentration of about 15 to 17$\,$ppm. To generate vacancies, electron irradiation was carried out with 1$\,$MeV and a dose of 3$\cdot 10^{18}\,$cm$^{-2}$. Afterwards, the diamond was annealed for 2 hours at 950 degrees and 10$^{-5}\,$mbar resulting in a NV concentration of approximately 1$\,$ppm. The final layer thickness was estimated from the CVD growth parameters to be around 400$\,$nm.

Decreasing the lateral size of the diamond sensor proved to be a crucial step for enabling measurement reproducibility. The diamond was laser cut to a size of 0.6 x 0.6 x 0.5 mm³ from its initial dimensions of 4x4x0.5 mm³. This corresponds to an over 40 fold reduction of the contact area between diamond and sample. Therefore, it is much less likely to catch a surface unevenness or a small particle when approaching the diamond to the sample. Also, slight unevenness becomes negligible due to the small contact area.

The collection optics and the camera sensor also have a strong influence on the measurement performance. The higher the NA of the objective, the more fluorescence light can be collected increasing the signal-to-noise ratio of the measurement. Furthermore, the quantum efficiency of the camera sensor as well as the read-out noise and dark noise determine the photon collection efficiency. 

The material samples need to be very close to the NV layer to achieve high spatial image definition of their magnetic features. As a rule of thumb, the maximum achievable resolution is in the order of the standoff between sample and NV layer. Therefore, the standoff should be smaller than 1$\,$µm for maximum measurement performance. To align the material samples very close to the NV layer, the surfaces of both sample and diamond have to be very flat.

\begin{figure}
\includegraphics[width=1\linewidth]{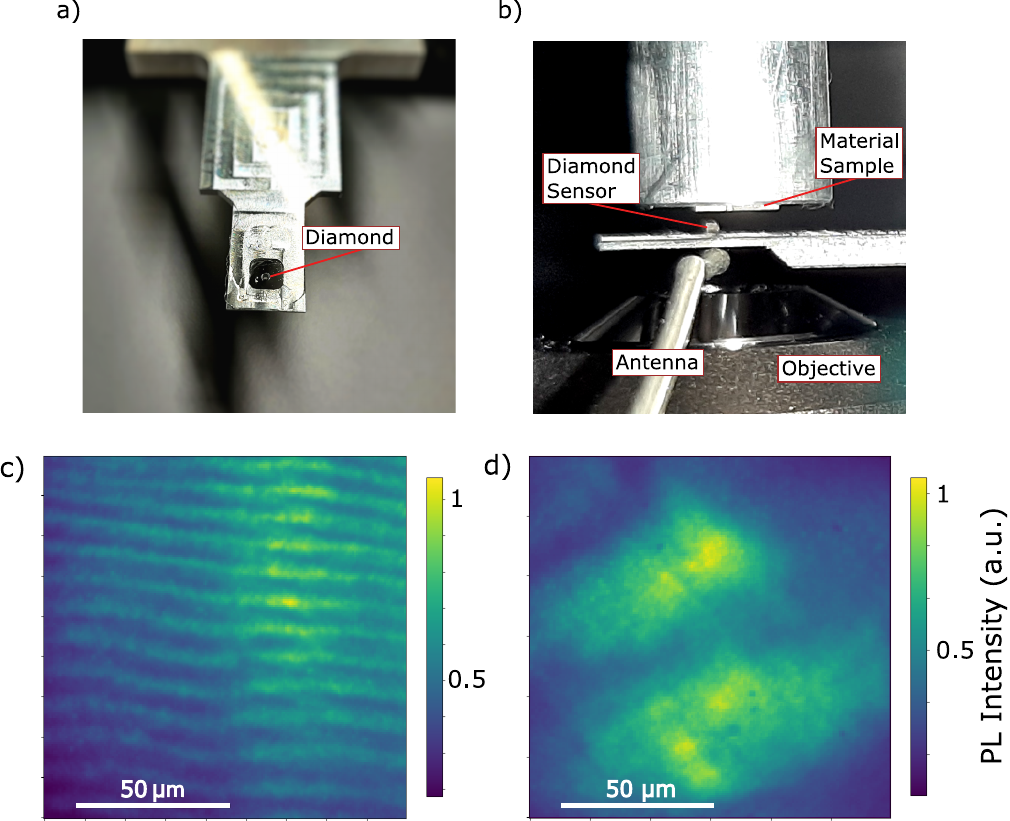} 
\centering
\caption{Photograph of a) the diamond sensor fixed on a custom holder without sample and b) the diamond brought close to a material sample glued on an aluminum rod (side view). Both diamond and sample can be positioned individually, the diamond position angles can also be altered for alignment optimization. Interference pattern formed between diamond and sample surface during alignment c) far from optimum alignment and d) after alignment optimization.}
\label{fig:photograph} 
\end{figure}

Fig. \ref{fig:photograph}a shows the diamond which is fixed on a custom holder. 
 The diamond is glued on top of a microscope glass slide using transparent optical adhesive. All positioning angles of the diamond can be altered using a stage system. The samples are glued on top of aluminum rods (see Fig. \ref{fig:photograph}b) which can be fixed on a micro stage enabling to alter the sample position with sub-micron accuracy. For optimizing the alignment, the interference pattern generated by the laser light reflection from the sample surface and the diamond surface is observed (see Fig. \ref{fig:photograph}c,d). A large angle between the two surface leads to small features in the interference pattern. Adjusting the angle of the diamond with respect to the sample surface, the pattern feature size is maximized. It was not possible to reach an alignment in which no interference pattern was visible. Nevertheless, a result like the one presented in Fig. \ref{fig:photograph}d proved to be sufficient for reproducible high spatial resolution ODMR measurements. 

\subsection{Simulations of the measurement process}
\begin{figure*}
\includegraphics[width=1\linewidth]{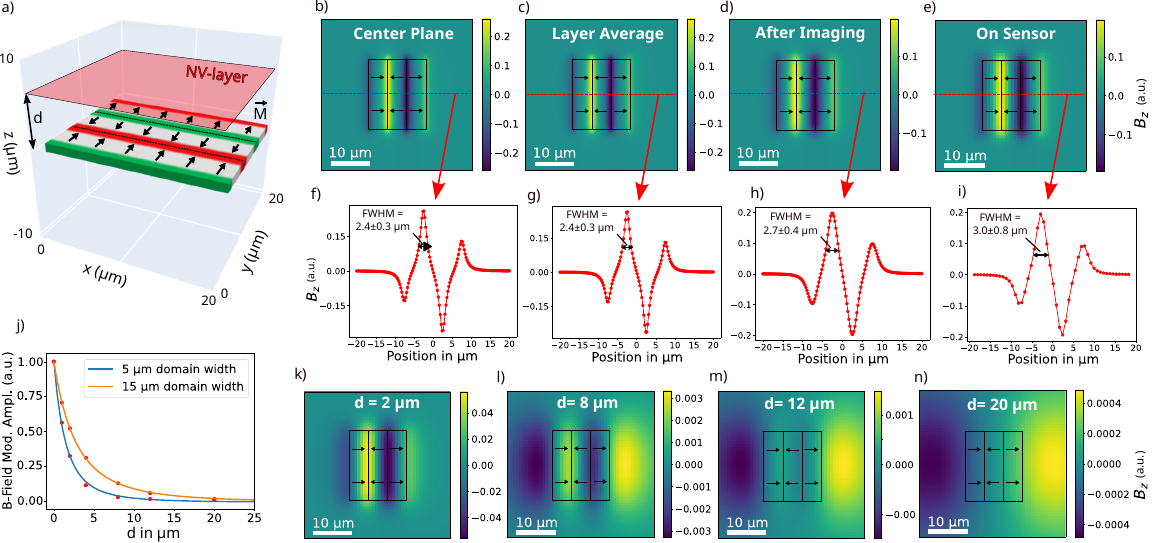} 
\centering
\caption{\label{fig:simulation_full} a) 3D illustration of the simulated magnetic domain pattern. The size of the domains is 20 x 5 x 1 µm³ and the magnetic polarization is arbitrarily set to 1$\,$T. b-i) Simulation of the NV-center based magnetic field imaging of the simulated pattern. The z magnetic field value is presented first only in the center plane of the NV layer before the layer thickness is taken into account by averaging over the respective plane positions. The imaging process is taken into account by a Gaussian convolution with a width determined by the optical resolution limit. The resulting magnetic field distribution is finally binned to simulate the effect of the sensor pixels. For each step, the cross section of the magnetic field distribution along the indicated slashed red line is plotted in the images f-i. j) Relation of the magnetic field modulation amplitude and the distance from the sample surface for the 5$\,$µm domain width pattern and for a second, similar pattern with 15$\,$µm domain width. Both curves show a strong decrease of the amplitude approximately proportional to 1/$r^3$ while the decay is less pronounced for the wider pattern size. k-n) Influence of the distance between NV layer and domain structure on the captured magnetic field map. The simulations were carried out based on the python package Magpylib \cite{Magpylib}. }
\end{figure*}

In order to assess the capability of widefield NV center magnetometry to detect domain distributions, simulations of both the magnetic field distribution and the measurement process were carried out \cite{Magpylib}. The magnetic stray field generated by an alternating domain pattern was simulated (see Fig. \ref{fig:simulation_full}a). The size of the domains is 20 x 5 x 1$\,$$\mu$m³ and the magnetic polarization is set to 1$\,$T. First, the simulation of the magnetic field imaging process is carried out in several steps presented in Fig. \ref{fig:simulation_full}b-e. An orientation of the NV centers perpendicular to the layer surface was assumed. As a result, only the magnetic field component perpendicular to the sample surface is taken into account. In \ref{fig:simulation_full}b the magnetic field in the center plane of the NV layer is presented. A realistic alignment was supposed to leave a 500$\,$nm residual standoff between diamond and sample. The NV layer position and thickness (a thickness of 400$\,$nm is assumed) was then taken into account by averaging the magnetic field over the NV layer thickness in \ref{fig:simulation_full}c. The optical imaging on the camera sensor was simulated in \ref{fig:simulation_full}c by convolution with a 2D Gaussian function of parameters determined by the optical diffraction limit (see equation \ref{eq:resolution}). The resulting magnetic field map was binned into squares corresponding to the pixel size of the camera sensor in Fig. \ref{fig:simulation_full}d. The pattern exhibits strong peaks at the domain borders. While the NV layer thickness does not affect the map at all, the imaging process and the limited pixel size both slightly blur the magnetic field image. This behavior was verified by calculating the horizontal width of the magnetic field pattern peak width for the maps corresponding to the different simulation steps (Fig. \ref{fig:simulation_full} f-i).

The distance between the NV layer and the magnetic domains strongly affects the magnetic field distribution at the position of the NV layer. The magnetic field modulation amplitude (see Fig. \ref{fig:simulation_full}j) reduces by a factor of approximately 1/r³, as follows from the simulation for two different alternating magnetic domain patterns (5$\,\mu$m and 15$\,\mu$m domain width, see  \ref{fig:simulation_full}j). The magnetic field modulation amplitude is defined as the difference between the minimum and the maximum magnetic field in the NV layer and decreases even faster for small domain patterns compared to larger ones. The spatial resolution (see Fig. \ref{fig:simulation_full}k-n) also drastically reduces when increasing the distance of the NV layer from the domain pattern.  As a consequence, the magnetic field must be measured very close to the sample surface to be able to draw conclusions about microscopic changes in the material. The simulations also show that, if the magnetic field is sensed very close to the material surface, information from the first few microns beneath the sample surface can potentially be acquired assuming typical domain sizes to be in the order of 5 to 15 $\,\mu$m. Larger domains can be detected from deeper inside the sample while smaller domains can only be resolved when they are close to the sample surface.

\subsection{Influence of the NV-orientation on the measurement}
Magnetic stray field maps recorded from an electrical steel sample were measured using different crystal-cut diamond orientations and therefore different NV orientations with respect to the targeted samples. While NV layers CVD grown on (100) diamond surfaces usually contain all four orientations of NV centers, it is also possible to achieve layers containing NV centers preferentially oriented perpendicular to the surface by overgrowing a (111) oriented diamond surface. The (111) diamond plate (3 × 3 mm$^2$, 0.3 mm thick, IIa HPHT-grown) was homoepitaxially overgrown at Fraunhofer IAF by microwave plasma-assisted chemical vapour deposition (MPCVD) in order to produce a well-defined, homogeneous NV-containing layer near the diamond surface whose properties are largely decoupled from those of the substrate. CVD growth processes were conducted using a 915 MHz ellipsoidal reactor. A 5 $\mu$m thick intrinsic buffer layer was deposited prior to the N-doped layer, which was subsequently overgrown at an N/C ratio of 5000 ppm and a CH$_4$/H$_2$ concentration of 0.5 $\%$ in the same process, the latter chosen to enforce a step-flow growth mode. This resulted in an N-doped layer with a P1 density of about 5–10 ppm, a thickness of roughly 300 nm, and preferentially oriented NV centres perpendicular to the surface in the as-grown layer. After growth and prior to the measurements, the diamond was exposed to an O$_2$ plasma (300 W, 15 min) to obtain a clean, oxygen-terminated surface, which stabilizes the negative charge state (NV$^-$) of near-surface NV centres.

\begin{figure}
\includegraphics[width=1\linewidth]{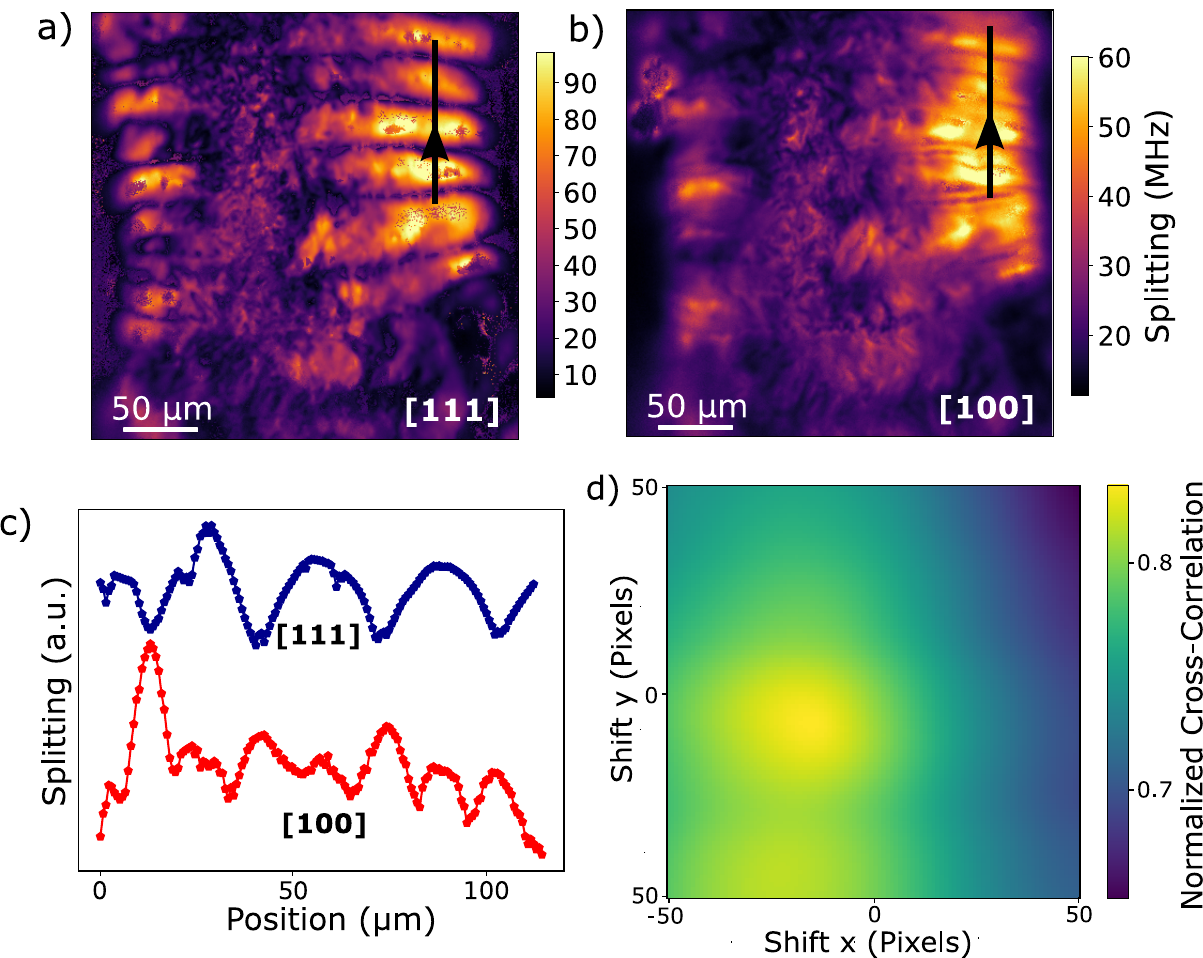} 
\centering
\caption{\label{fig:111_experiment} Zeeman splitting map measured from a fatigued electrical steel sample recorded using a) a (111) diamond plate containing only one NV orientation perpendicular to the diamond surface and b) a (100) diamond plate containing all four NV orientations. c) Comparison of the measurements along the lines indicated in a and b. d) Cross-correlation of the measurements a and b. }
\end{figure}

The interpretation of a measurement recorded based on a (111) diamond is much simpler to interpret. Due to the preferential orientation, only the magnetic stray field component perpendicular to the diamond surface (and therefore parallel to the NV axes) induces a Zeeman splitting. When using a (100) diamond with a layer containing all four NV orientations, the resonances can in general not be clearly distinguished without using a sufficiently strong magnetic bias field separating the respective resonances. In the absence of such a bias field, the resonances overlap, in many cases resulting in two relatively broad resonances. Fitting a double Lorentzian to these two peaks allows to determine a value which is close to the average splitting of all four peaks. Due to the orientation of the NV center within the diamond crystal lattice, the determined splitting value is correlated with the magnetic field component perpendicular to the surface. Nevertheless, the overlap of the resonances corresponding to the four orientations induces artifacts in the magnetic field map. A direct comparison of measurements recorded from the same electrical steel sample at the same position using both diamond types was carried out and is presented in Fig. \ref{fig:111_experiment}. 

While the main features captured by the (111) diamond are also visible in the (100) measurement, additional features can be observed while the imaging appears to be generally more washed out. Both effects occur due to the four different NV orientations and the subsequent overlap of the resonances. There is always a magnetic field component aligned to one of the NV orientations. When the magnetic field orientation rotates, several extrema of the measured average Zeeman splitting can be expected to appear due to the change of the field components parallel to the different NV orientations. Line plots were compared for the two measurements at the same position (see Fig. \ref{fig:111_experiment}c). The (111) measurement shows well defined minima and maxima. The (100) measurement exhibits a maximum for all minima of the (111) measurement, but additionally there is always one maximum coinciding with the maxima in the (111) measurement. The cross correlation of the measurements presented in Fig. \ref{fig:111_experiment}d shows a broadened peak indicating a correlation of the images. Since the diamond and the sample were realigned in between the measurements, the peak might be washed out due to slight changes of the sample position. Also, the artifacts appearing in the (100) measurement reduce the correlation.

\begin{figure}
\includegraphics[width=1\linewidth]{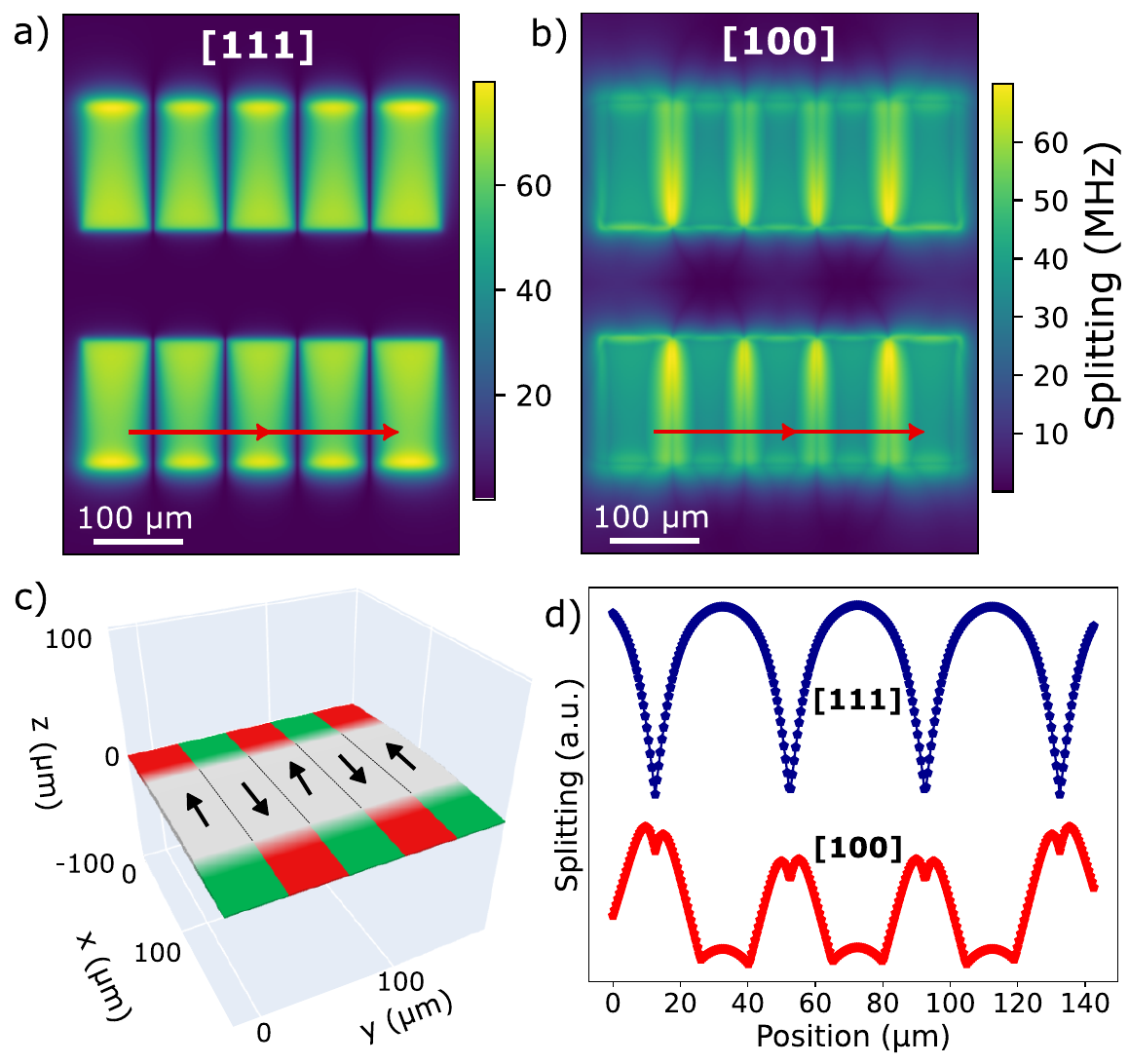} 
\centering
\caption{\label{fig:111_simulation} Simulated splitting maps using Magpylib \cite{Magpylib} for a) a (111) diamond containing only one NV orientation perpendicular to the surface and b) a (100) diamond containing all four NV orientations. c) 3D illustration of the magnetic domain orientation used for the simulation. The magnetic domains are assumed to be oriented in the material sample plane. The shape of the sample was taken into account by reducing the thickness of the structure towards the sides. d) Comparison of the two simulations along the indicated lines. }
\end{figure}

To better understand the measurements captured for the two diamond types, simulations were carried out (see Fig. \ref{fig:111_simulation}). An alternating magnetic surface domain pattern, as it is expected for the electrical steel sample to form, was simulated and the magnetic stray field distribution was calculated in the plane containing the NV centers. The top surface of the simulated domains is chosen to slightly recess towards the sides as it is the case for the real sample. The magnetic domain structure is expected to be formed by alternating domains oriented in the plane of the material sample. The simplified Hamiltonian was solved by calculating the magnetic field projection on the respective NV axis and extracting the energy levels. For the (111) diamond, the simulation almost perfectly matches the experimentally determined Zeeman splitting distribution. Due to the in-plane orientation of the domain magnetization, the z magnetic field component vanishes in the center. The simulation for the (100) diamond on the other hand only partially fits the experimental measurement. This could be due to deviations of the simulated domain distribution in comparison with the real sample, but also due to additional superposing magnetic stray fields generated deeper below the sample surface. Nevertheless, the simulation helps to roughly understand the differences between the measurements based on a (100) and a (111) diamond. While measuring with a (111) diamond sensor seems to be the superior choice due to the well defined stray field features and a simpler interpretation, these diamonds usually contain much less NV centers compared with NV layers grown on (100) diamonds. Therefore, the fluorescence signal is much lower and the measurement duration vastly increases using a (111) diamond. CVD layers grown on (111) diamond usually exhibit a much higher surface roughness making the alignment of the diamond with respect to the sample much harder. Since the main features of the magnetic stray field maps are contained in both measurements, using a (100) diamond sample proved to be a much more efficient choice for non-destructive testing applications. For this reason, all measurements presented as follows are based on a (100) diamond sensor.

\section{Results and Discussion}
\subsection{General Considerations and Measurement Interpretation}
To evaluate the applicability of widefield NV center magnetic imaging for non-destructive testing applications, measurements were carried out on different types of material. It was found that reproducible measurements can be carried out for soft magnetic materials. Hard magnetic materials generally generate strong magnetic stray fields. The latter induce a quenching of the NV center fluorescence making it extremely difficult to achieve a well-defined magnetic map. Soft magnetic ferromagnetic samples on the other hand are strongly magnetized within external magnetic fields leading to similar challenges. Therefore, measurements were carried out in the absence of a magnetic bias field. Only a small magnetic bias field of about 100$\,$µT was present during the measurements generated by the earth magnetic field and the lab surrounding. As a result, the resonances corresponding to the four NV orientations overlapped in most measurements. The estimated Zeeman splitting value therefore corresponds to the average splitting of the four NV orientations. The maps are generated by fitting a double Lorentzian to the ODMR spectrum acquired for each pixel.

The measured stray field corresponds to the magnetic field outside the material resulting from its internal magnetization distribution and associated demagnetizing field.
Since magnetic flux tends to close near the surface, the stray field decays rapidly with increasing distance from the sample. Consequently, magnetic imaging is most sensitive to
near-surface features, while signals originating from deeper structures are strongly attenuated. The achievable spatial resolution therefore depends critically on the sensor–sample distance, with high spatial frequency components decaying most rapidly.

\begin{figure}
\includegraphics[width=1\linewidth]{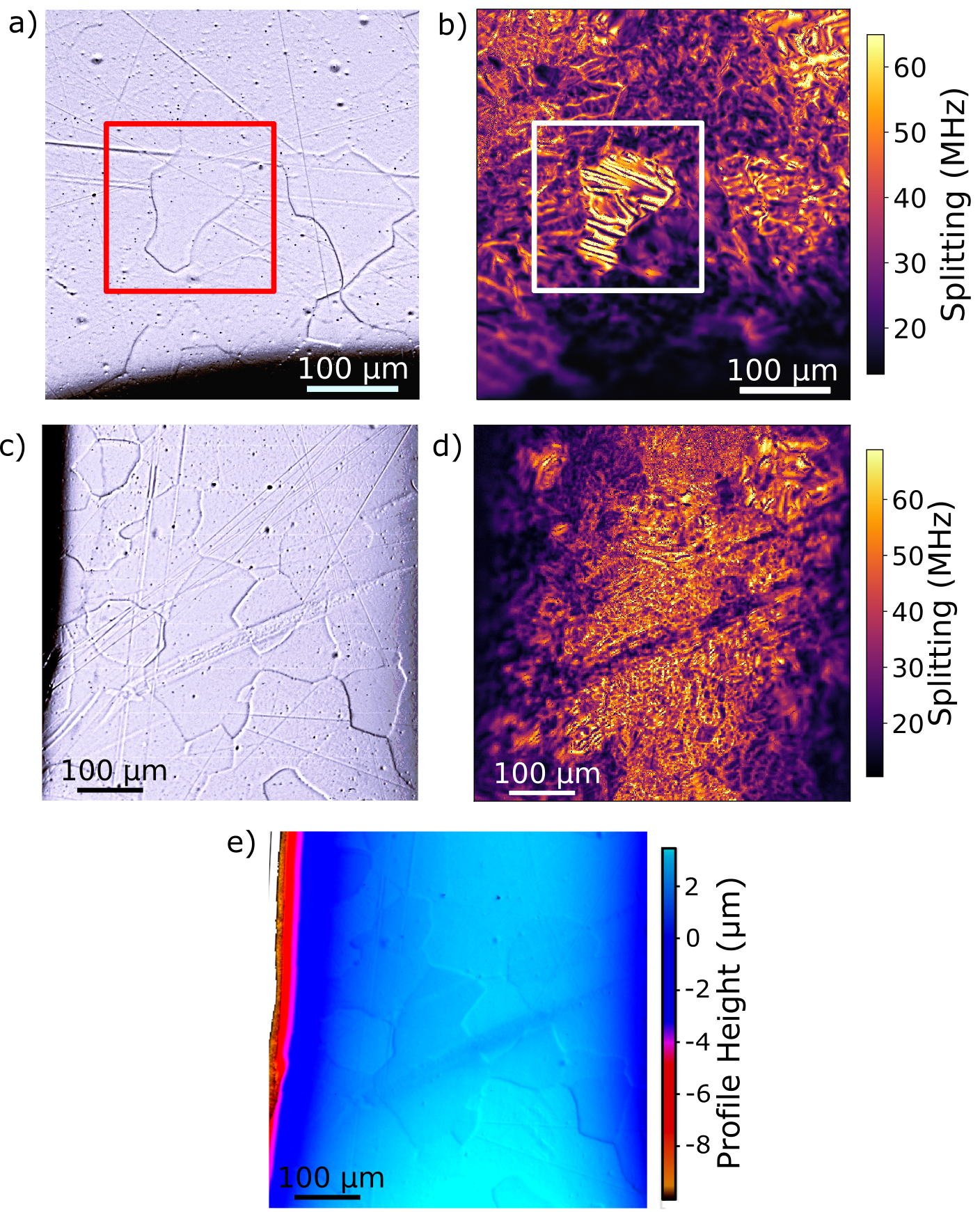} 
\centering
\caption{\label{fig:grain_boundary}a) Light microscopy image and b) Zeeman splitting map recorded from a pristine iron sample exhibiting several grains. c) Light microscopy image, d) Zeeman splitting map and e) height profile (measured using a laser profiler) of a pristine iron sample. Only the most prominent scratch seems to influence the magnetic stray field pattern. }
\end{figure}

Fig. \ref{fig:grain_boundary}b shows a measurement carried out on a polished pure iron sample possessing multiple grains in the field of view. Comparing the magnetic stray field map measured by NV magnetometry with the light microscopy image in \ref{fig:grain_boundary}a, a strong correlation of grains and magnetic features can be observed. The effect is not the same for all grains which is most likely due to differing grain orientation. The highlighted grain exhibits prominent line-shaped patterns which are sharply limited by the grain boundary. The other grain boundaries have a less obvious effect on the magnetic stray field map, but there are still some visible differences of the measured stray field pattern shapes, and amplitudes. The different quality of the magnetic stray field patterns generated by the domains can be explained by a different orientation of the magnetic easy axis with respect to the sample surface. If there is a stronger mismatch, smaller patterns tend to form to reduce the total magnetic energy. The strong deviation of the measured magnetic field patterns from line-shaped patterns could be caused by residual stress induced by the polishing of the sample. The surface imperfections clearly visible in the microscope image only faintly influence the magnetic stray field map confirming the stray field is mainly influenced by the magnetic microstructure.
Fig. \ref{fig:grain_boundary}c-e shows a measurement taken from an iron sample with a prominent scratch and smaller side scratches within the field of view. Comparing the height profile of the sample to the Zeeman splitting map, it becomes apparent that only the most pronounced scratch strongly affects the NV measurement while the smaller scratches appearing in the light microscopy image do not appear at all. It can therefore be concluded that minor surface inhomogeneities do not strongly affect the NV measurement. Only prominent features strongly modulating the sample surface affect the NV measurement considerably. A theoretical examination of magnetic signals generated by surface defects can be found in \cite{Korner}.

\begin{figure}
\includegraphics[width=1\linewidth]{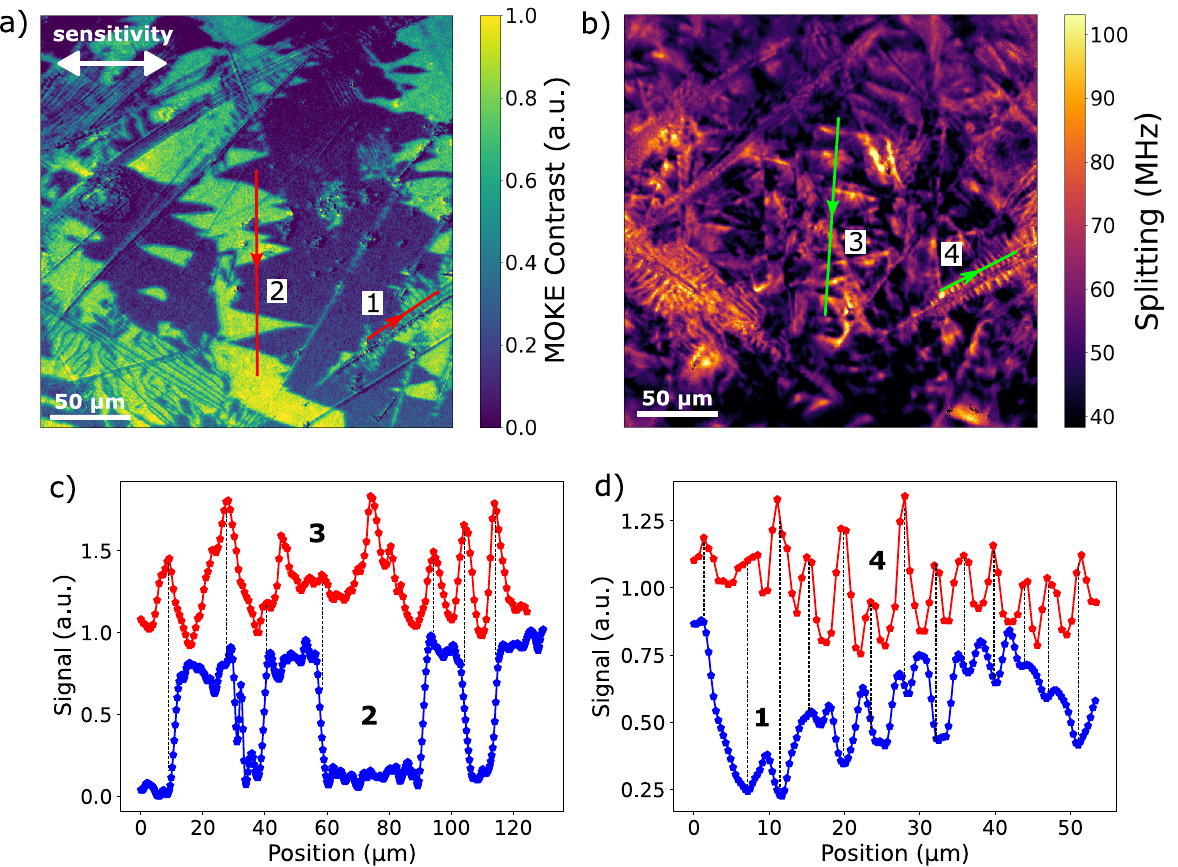} 
\centering
\caption{\label{fig:MOKE_line} a) longitudinal MOKE measurement and b) Zeeman splitting map recorded from an electrical steel sample. c,d) comparison of the measurements along the lines indicated in a,b).}
\end{figure}

To gain a better understanding of the relation between the magnetic microstructure and the magnetic stray field maps measured by NV magnetometry, a direct comparison to a MOKE measurement was carried out. While MOKE is only able to determine the surface magnetization, NV center magnetometry directly measures the magnetic field value. Correlations of both measurements are expected since the magnetic stray field is strongly influenced by the surface magnetization. But the magnetic field map is also expected to be influenced by magnetic fields generated deeper inside the material enabling to acquire additional information about the micro magnetic structure. Fig. \ref{fig:MOKE_line} shows a direct comparison of the two measurement techniques. A direct comparison of NV center magnetic field imaging and MOKE for a magnetic thin film can also be found in \cite{moke_nv_paper}. 
For the MOKE measurement, the light and dark areas indicate the relative orientation of the surface magnetization. A longitudinal MOKE contrast along the indicated sensitivity axis is used. The measurement data was compared along two lines, one of them positioned in the center area exhibiting large magnetic surface domains. The other inspection line position was chosen to be located at a position exhibiting a small and well-defined domain pattern. The pattern is most likely caused by stress concentrations on a large scale defect, such as a scratch. The MOKE measurement exhibits well-defined plateaus along the large magnetic domains indicating a well-defined orientation of the surface magnetization within the magnetic domains. The NV measurement on the other hand exhibits peaks which are  strongly correlated to the domain walls. This is expected since a Bloch wall involves the rotation of the magnetization vector out of the surface plane inducing a locally increased stray field. Also, the magnetic field lines can generally be expected to exhibit an increased density near the domain borders if the magnetization is not perfectly parallel to the sample surface (which is usually the case for any real material sample). Interestingly, there are also peaks visible in the NV measurement located far from the domain borders. Those peaks can be expected to be generated by buried magnetic domains and prove the capability of NV magnetometry to capture additional information that is not accessible using MOKE. 

\subsection{Measurements on Cyclically Loaded Samples}
\begin{figure*}
\includegraphics[width=1\linewidth]{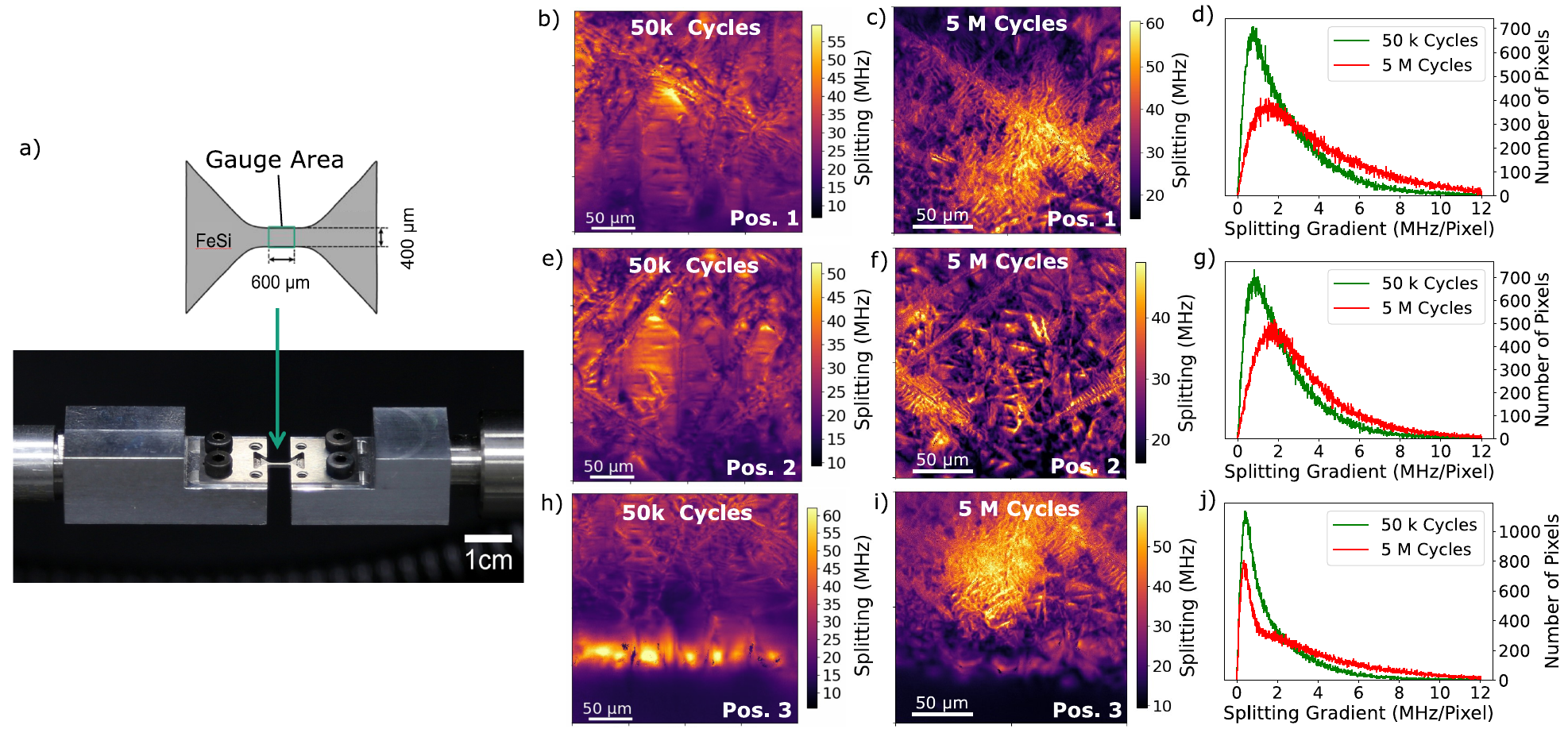} 
\centering
\caption{\label{fig:before_after} a) Sketch of the electrical steel sample and photograph of the configuration used for cyclic loading. The sample was clamped at the wide side parts, the Zeeman splitting maps were recorded at different positions within the gauge area. b-j) Comparison of the measured magnetic field maps generated by the electrical steel sample before and after cyclic loading for three different measurement positions. There is an obvious change in quality and quantity of the detected magnetic features. For each position, histograms of the splitting gradient are presented. The gradient distribution characteristically changes after fatigue shifting to a higher abundance of strong splitting gradients.}
\end{figure*}

Electrical steel is a very soft magnetic material known to form well defined magnetic domain patterns. Therefore, it is suitable for exploring the effect of material fatigue on the magnetic stray field distribution. Grain oriented electrical steel samples were surface prepared and cyclically loaded with well controlled forces. The samples were cut from a sheet of 0.28 mm thickness. The large grain size allowed to cut samples that do not contain any grain boundaries within the field of view. The samples were cut in a bone like shape offering large sides to clamp the sample and a narrow gauge area (400 x 600 µm²) in the center of the sample (see Fig. \ref{fig:before_after}a). The chemical composition of the electrical steel sample was 96.27$\,\%$ iron, 3$\,\%$ silicon, 0.7$\,\%$ manganese and 0.03$\,\%$ carbon. To ensure a planar surface and to remove corrosion, several grinding and polishing steps were carried out using a Struers Tegramin-30 polishing machine. To remove surface rust and slight pitting, the specimens were ground with 2500 grit sandpaper. Afterwards, polishing cloths with 9$\,$µm, 3$\,$µm, and 1$\,$µm diamond suspension were used to smoothen the surface. A final oxide polishing step (OPS) with 0.1$\,$µm colloidal silica suspension was used to achieve the final surface finish. The cyclic loading was carried out stepwise and a widefield NV center Zeeman splitting image corresponding to the magnetic field distribution was measured for several well-defined measurement positions after each loading cycle. The specimens were subjected to stress-controlled cyclic loading with a sinusoidal waveform. The stress amplitude was set to 120$\,$MPa at a stress ratio of R = 0.1 and a frequency of 60$\,$Hz. Since the maximum applied stress remained below the material's yield strength, the specimens operated in the high-cycle fatigue regime without macroscopic plastic deformation. This ensured that the specimen surface remained planar throughout the fatigue process, which is essential for maintaining optimal contact between the NV-diamond sensor and the sample surface during magnetic field measurements. After loading, the sample shape and surface did not exhibit any strong changes in microscopic inspection.

Fig. \ref{fig:before_after} shows a direct comparison of Zeeman splitting maps recorded after $5\times10^4$ and after $5\times10^6$ cyclical loading steps for different positions inside the gauge area of the sample. To release pre-existing strain and inhomogeneities in the sample to a certain extent, the initial measurements are taken after a low number of 50e3 cycles and are then compared to the measurement taken for a high number of 5e6 cycles. A clear change of quality and quantity of the magnetic stray field features can be observed for all three positions. After cyclical loading, the magnetic maps contain sharper defined features while the quantity of small features strongly increased. Some of the small features are oriented perpendicular to pre-existing large line-shaped features, most likely fine scratches. This observation can be explained by local stress concentrations forming during the cyclic loading due to the discontinuity of the sample surface.

The measurements were quantitatively evaluated by determination of the splitting gradient distribution. While the splitting maps are very different for all measurement positions, the splitting gradient distribution proved to characteristically shift after the cyclic loading. While most pixels of the gradient map initially have values below 2$\,$MHz/pixel, the distribution is shifted towards high splitting gradients between 4 to 10$\,$MHz/ pixel. This behavior offers the potential for a non-destructive, local detection of fatigue processes inside the sample. Even though clear changes of the splitting gradient distribution were observed, no damage or deformation of the sample was visible in light microscopy inspection. Therefore, the observed effects are suspected to be caused by an increase of microscopic defects introduced by cyclic loading. The defects cause local stress concentrations which induce a local alternation of the magnetic domain distribution. As a consequence, finer domain structures develop resulting in a higher abundance of strong stray field gradients in the measured magnetic stray field maps. While the shift of the splitting gradient distribution is very similar for the positions 1 and 2, the histograms are different for position 3 as there are much more pixels with a low splitting gradient. This is due to the sample boarder positioned within the field of view. Nevertheless, the histogram is shifted towards higher splitting gradients after cyclic loading, just as it is the case for the other measurement positions.

\begin{figure}
\includegraphics[width=1\linewidth]{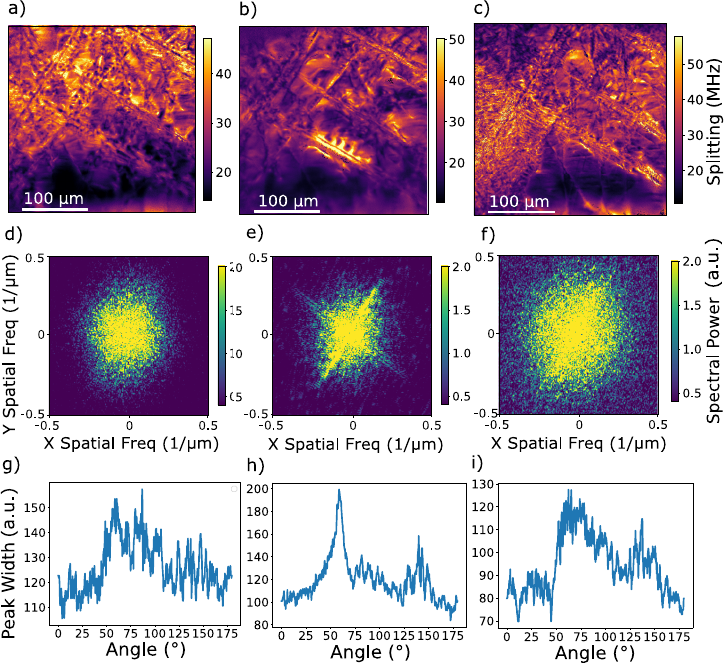} 
\centering
\caption{\label{fig:fatigue_peak} Measurements of the magnetic stray field maps on an electrical steel sample carried out during step-wise cyclic loading of the sample a) before the loading, b) after 500e3 steps and c) after 2e6 steps. The respective 2D Fourier transform plots are presented in d-f). g-i) show the width over angle plots for each step. Plotting the width of the 2D Fourier transform peak over the respective angle, a strong peak appears after 500e3 cycles together with a smaller side peak. After 2 million loading cycles, the peaks mostly disappear again while the high-frequency signal contribution strongly increases.}
\end{figure}

An additional method for evaluating the captured Zeeman splitting maps is based on 2D Fourier transformation. This approach is mainly suited to detect strong magnetic features that can develop during the fatigue process. As these features are distributed statistically, the obtained results strongly differed depending on the position on the sample. Fig. \ref{fig:fatigue_peak} shows the results for one of the measurement positions.  The Zeeman splitting maps together with the respective 2D Fourier transforms and the Fourier peak with over angle plots generated from the 2D Fourier transforms measured before the loading, after 100e3 cycles, after 500e3 cycles and after 2 M cycles are presented. Before the loading, the map contains a large amount of relatively small magnetic features with rather low amplitudes together with a few line-shaped features, most likely fine scratches in the material surface. After 200e3 loading steps, the quantity of small features reduced while large features with strong amplitudes start to form. The shape of the corresponding 2D Fourier transform is strongly distorted along one direction and a peak started to form subsequently in the width over angle plot. After 500e3 loading steps, a very prominent magnetic structure developed along one of the pre-existing line shaped features. The 2D Fourier transform is strongly distorted along two directions, a sharp and well-defined peak together with a smaller peak are visible in the width over angle plot. The deviation of the 2D Fourier transform from a round shape is supposedly caused by directed magnetic features forming along the pre-existing discontinuities. After 2 Million loading cycles, the 2D Fourier transform almost took a circular shape again, but the general width of the peak strongly increased. This is caused by a large quantity of fine magnetic features that evolved in the Zeeman splitting map. This observation underlines the claim made by evaluating the splitting gradient distributions. A larger width of the 2D Fourier transform peak corresponds to higher frequencies present in the splitting map and therefore also to an increase of high splitting gradients. The large scale magnetic features that were observed before almost disappeared completely. The former two peaks are barely visible in the width over angle plot. As mentioned before, this behavior is supposedly caused by a large amount of defects introduced by the cyclic loading leading to stress concentrations and pinning of domain walls and dislocations. While data evaluation via 2D-Fourier transform can give valuable additional insight about local processes taking place inside the sample during the fatigue process, the approach based on evaluating the splitting gradient proved to be overall more consistent and reproducible for non-destructive testing. 

\subsection{Optimizing Measurement Time}
\begin{figure}
\includegraphics[width=0.9\linewidth]{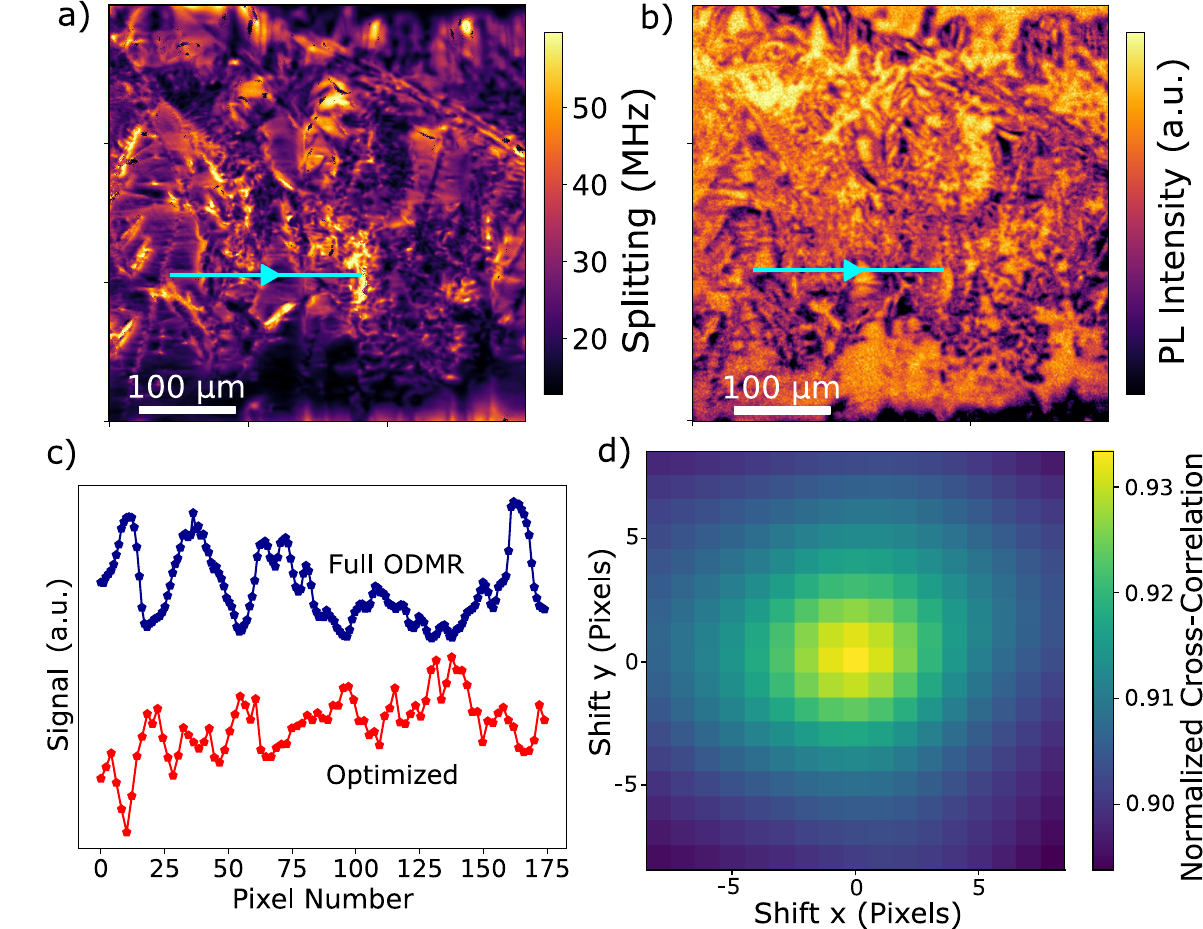} 
\centering
\caption{\label{fig:9s} a) Zeeman splitting map recorded from a fatigued electrical steel sample. b) Map of the PL intensity measured based on a microwave frequency fixed at the highest slope position of the resonance. c) Line comparison of the two measurements. d) Cross correlation of the two measurements.}
\end{figure}

Non-destructive testing generally requires fast and reproducible measurement techniques. To reduce the measurement time, it can be very beneficial not to measure the whole ODMR spectrum but to fix the microwave frequency at the highest slope position of one of the resonances (see {Fig.}\ref{fig:energy_levels}b). While the resulting measurement is more effortful to evaluate quantitatively and might also lose some features due to the limited width of the ODMR peak, it still proved to contain most features that were also measured by the full ODMR protocol. Fig. \ref{fig:9s} shows a direct comparison of the full ODMR measurement and a second measurement based on a fixed microwave frequency at the highest slope position of the resonance. While the latter contains more noise and is also inverted compared to the full ODMR measurement (depending on which side of the resonance peak is chosen), the cross correlation of the two images is very high and well defined. While the full ODMR measurement required approximately 5 minutes to record, the fixed-frequency measurement could be recorded in only 9 seconds. By further improving the setup, measurement times below one second can be expected to be feasible. Tailoring the NV layer thickness to the achievable spatial resolution (1-2 µm) can be expected to increase the measurement signal allowing a further reduction of the measurement time. Also, optimizing the fluorescence light collection efficiency would be possible using a high numerical aperture, as provided by, for instance, oil immersion microscope objectives. The latter step requires a reduction of the diamond thickness as well as a lithographic antenna due to the limited working distance of immersion microscope objectives. Coupling the laser beam into the diamond from the side would additionally allow to vastly increase the excitation power to increase the signal even more.

\section{Conclusion}
This work demonstrates the applicability of widefield NV center magnetometry for non-destructive testing applications. The measurement approach allows to capture high spatial resolution Zeeman splitting maps directly related to the magnetic stray field distribution emitted by the sample. The measurement is carried out based on a layer containing a high density of NV centers near the surface of a diamond plate. Compared to traditional scanning probe approaches, this technique is fast, robust and provides a relatively large field of view. Characteristic changes of the captured magnetic Zeeman splitting distributions could be shown for an electrical steel sample after cyclic loading. 
After 5e6 fatigue cycles, a vast increase of tiny magnetic features could be observed for all measurement positions. This behavior was quantified based on the Zeeman splitting gradient distribution calculated from the measurements via histograms. The cyclic loading shifts the histogram towards high splitting gradients. This observation is consistent for all measurement positions on the sample. The characteristic shift of the splitting gradient distribution offers the potential to be used as a marker for early fatigue damage, even before cracks and large discontinuities are introduced to the sample. The increasing number of small magnetic anomalies can presumably be attributed to  microscopic material defects introduced by the cyclic loading. The defects induce local strain and lead to pinning of domain walls altering the magnetic microstructure. An additional method for evaluating the measurements based on 2D Fourier transforms was explored offering a way to detect directed magnetic features formed during the fatigue process. Test measurements confirmed that the Zeeman splitting maps are mainly determined by the magnetic microstructure of the material while surface effects play a subordinate role. It could be shown that buried magnetic features can potentially be detected from at least the top few microns below the sample surface, which is not possible using established techniques like MOKE. Nevertheless, a direct reconstruction of the magnetic domain distribution from a 2D magnetic field map is generally not possible. Due to the variety of contributions locally influencing the magnetic stray field of a material sample, further simulations and test measurements have to be carried out for a better understanding of the captured data.

NV center magnetometry is mainly suited as a non-destructive testing method for miniaturized material samples. Tests on small samples taken from critical areas of larger parts can provide valuable information on fatigue mechanisms in these components. Additionally, gaining a better understanding of material fatigue on small scales can contribute to the further optimization of materials. Furthermore, material fatigue is still not well understood on a microscopic scale, motivating detailed monitoring of the processes involved in early-stage fatigue.

\section*{Author Contributions}
NM, XV designed the quantum sensing experiments and the instrument. MF, SP and TS picked and prepared the material samples. PK grew the NV doped diamond. NM carried out the measurements and simulations and developed data evaluation methods. MF prepared the material samples and carried out the sample fatigue. IS performed the MOKE measurements. NM, MF, SP, IS, TS, JJ, RQ and XV analyzed and validated the data and discussed the results. NM and XV wrote the original draft. All authors participated in discussion and editing of the manuscript.

\section*{Declaration of competing interest}
There are no conflicts to declare.

\section*{Funding}
This research was funded by the Fraunhofer Lighthouse Project Quantum Magnetometry (QMag) and the Fraunhofer Project NextLevel Quantum Magnetometry. This work has been partially funded by The European Union’s research and innovation programme through the PROMISE project (Grant Agreement No. 101189611).

\section*{Acknowledgments}
The experimental work was performed at Fraunhofer IAF.

\section*{Data availability}
The data supporting the findings of this study are available from the corresponding author upon reasonable request.

\bibliographystyle{unsrt}
\bibliography{cas-refs}

\end{document}